\documentclass[12pt]{article}
\usepackage[dvips]{graphicx}

\begin{document}

\author{Irina Radinschi$^1$\thanks{%
iradinsc@phys.tuiasi.ro} and I-Ching Yang$^2$\thanks{%
icyang@nttu.edu.tw} \and Department of Physics, ''Gh. Asachi''
Technical
University, \and Iasi, 6600, Romania \and and Department of
Natural Science
Education, \and National Taitung University, Taitung, Taiwan 950,
\and %
Republic of China}
\title{On the Energy of Stringy Black Holes }
\date{February 29, 2004}
\maketitle

\begin{abstract}
It is well-known that one of the most interesting and challenging
problems
of General Relativity is the energy and momentum localization.
There are
many attempts to evaluate the energy distribution in a general
relativistic
system. One of the methods used for the energy and momentum
localization is
the one which used the energy-momentum complexes. After the
Einstein work, a
large number of definitions for the energy distribution was
given. We
mention the expressions proposed by Landau and Lifshitz,
Papapetrou,
Bergmann, Weinberg and M\o ller. The Einstein, Landau and
Lifshitz,
Papapetrou, Bergmann and Weinberg energy-momentum complexes are
restricted
to calculate the energy distribution in quasi-Cartesian
coordinates. The
energy-momentum complex of M\o ller gives the possibility to make
the
calculations in any coordinate system.

In this paper we calculate the energy distribution of three
stringy black
hole solutions in the M\o ller prescription. The M\o ller
energy-momentum
complex gives us a consistent result for these three situations.


PACS: 04. 20 Dw, 04. 70. Bw,
\end{abstract}

\section{INTRODUCTION}

It is well-known that one of the most interesting and challenging
problems
of general relativity is the energy and momentum localization.
Numerous
attempts have been made in the past for a solution, and this
problem still
attracts considerable attention in the literature, and remains an
important
issue to be settled. The different attempts at constructing an
energy-momentum density don't give a generally accepted
expression.

Related on the method that used the energy-momentum complexes we
can say
that there are various energy-momentum complexes including those
of Einstein
[1]-[2], Landau and Lifshitz [3], Papapetrou [4], Bergmann [5],
Weinberg [6]
and M\o ller [7]. Also, there are doubts that these prescriptions
could give
acceptable results for a given space-time. The problem is that
with
different energy-momentum complexes we can obtain different
expressions for
the energy associated with a given space-time. This is because
most of the
energy-momentum complexes are restricted to the use of particular
coordinates. But the results obtained by several authors [8]-[9]
demonstrated that the energy-momentum complexes are good tools
for
evaluating the energy and momentum in general relativity.
Although, the
energy-momentum complexes of Einstein, Landau and Lifshitz,
Papapetrou,
Bergmann and Weinberg are coordinate dependent they can give a
reasonable
result if calculations are carried out in quasi-Cartesian
coordinates. Some
interesting results sustain this conclusion.

On the other hand, M\o ller [7] constructed an expression which
enables one
to calculate the energy distribution in any coordinate system not
only in
quasi-Cartesian coordinates. Also, Lessner in his important work
[10]
pointed out that the problem lies with the interpretation of the
result from
a special relativistic point of view instead of a general
relativistic one.
In conclusion, the M\o ller prescription can be used with success
to
evaluate the energy distribution of a given space-time. Many
results
recently obtained [11] support the conclusion given by Lessner in
his recent
paper that the M\o ller definition is a powerful concept of
energy and
momentum in general relativity.

In this paper we calculate the energy distribution of some
stringy black
hole solutions in the M\o ller prescription. We study the energy
associated
with these solutions because we think they can furnishes us
interesting
results. We have three cases. The first case is that of the dual
solution
known as the magnetic black hole solution. The metric is obtained
by
multiplying the electric metric in the Einstein frame by a factor
$%
e^{-2\,\Phi }$. In the second case we consider the metric which
describes a
non-asymptotically black hole solution in dilaton-Maxwell gravity
and was
given by Chan, Mann and Horne [13]. First, we perform the
calculations for
the string metric (see Kar [13]). In the last case we study the
string
metric for the magnetic black hole (Kar [13]). The M\o ller
energy-momentum
complex gives us consistent results for these three situations.
Through the
paper we use geometrized units ($G=1,c=1$) and follow the
convention that
Latin indices run from $0$ to $3$.

\section{ENERGY\ IN\ THE\ M\O LLER\ PRESCRIPTION}

The low energy effective theory largely resembles general
relativity with
some new ''matter'' fields as the dilaton, axion etc [13]-[14]. A
main
property of the low-energy theory is that there are two different
frames in
which the features of the space-time may look very different.
These two
frames are the Einstein frame and the string frame and they are
related to
each other by a conformal transformation ($g_{\mu \nu
}^E=e^{-2\,\Phi
}g_{\mu \nu }^S$) which involves the massless dilaton field as
the conformal
factor. The string ''sees'' the string metric. Many of the
important
symmetries of string theory also rely of the string frame or the
Einstein
frame [15].

The action for the Einstein-dilaton-Maxwell theory is given by

\begin{equation}
S_{EDM}=\int d^4x\sqrt{-g}e^{-2\,\Phi }[R+4\,g_{\mu \nu }\,\nabla
^\mu
\,\Phi \,\nabla ^\nu \Phi -\frac 12g^{\mu \lambda }g^{\nu \rho
}F_{\mu \nu
}F_{\lambda \rho }].  \label{1}
\end{equation}

Varying with respect to the metric, dilaton and Maxwell fields we
get the
field equations for the theory given as

\begin{equation}
R_{\mu \nu }=-2\,\nabla _\mu \,\Phi \,\nabla _\nu \Phi +2\,F_{\mu
\lambda
}F_\nu ^{\,\,\,\,\lambda }\,,  \label{2}
\end{equation}

\begin{equation}
\nabla ^\nu (e^{-2\,\Phi }F_{\mu \nu })=0,  \label{3}
\end{equation}

\begin{equation}
4\,\nabla ^2\Phi -4(\nabla \Phi )^2+R-F^2=0.  \label{4}
\end{equation}

The metric (in the string frame) which solve the
Einstein-dilaton-Maxwell
field equations to yield the electric black hole is given by

\begin{equation}
ds^2=A(1+\frac{2\,M\,\sinh ^2\alpha }r)^{-2}\,dt^2-\frac
1A\,dr^2-r^2\,d\theta ^2-r^2\,\sin ^2\theta \,d\varphi ^2. 
\label{5}
\end{equation}

where $A=1-\frac{2\,M}r$.

In the string frame the dual solution known as the magnetic black
hole is
obtained by multiplying the electric metric in the Einstein frame
by a
factor $e^{-2\,\Phi }$. Therefore, the magnetic black hole metric
is given by

\begin{equation}
ds^2=\frac AB\,dt^2-\frac 1{A\,B}\,dr^2-r^2\,d\theta ^2-r^2\,\sin
^2\theta
\,d\varphi ^2,  \label{6}
\end{equation}
with $B=1-\frac{Q^2}{M\,r}$.

Recently, some non-asymptotically flat black hole solutions in
dilaton-Maxwell gravity are due to Chan, Mann and Horne [12]. The
metric of
the solution in the Einstein frame is given by

\begin{equation}
ds^2=\frac C{\gamma ^4}\,dt^2-\frac 1C\,dr^2-r^2\,d\theta
^2-r^2\,\sin
^2\theta \,d\varphi ^2,  \label{7}
\end{equation}

where $C=r^2-4\,\gamma \,M$.

The string metric is obtained by performing the usual conformal
transformation on the metric. This turns out to be

\begin{equation}
ds^2=\frac{r^2\,D}{\gamma ^4}\,dt^2-\frac 1D\,dr^2-r^2\,d\theta
^2-r^2\,\sin
^2\theta \,d\varphi ^2,  \label{8}
\end{equation}

with $D=1-\frac{2\,\sqrt{2}\,\gamma ^2\,M}{Q\,r}$.

The string metric for the magnetic black hole is given by

\begin{equation}
ds^2=\frac{2\,Q^2\,E}{\gamma ^4}\,dt^2-\frac{2\,Q^2}{r^2\,E}%
\,dr^2-2\,Q^2\,r^2\,d\theta ^2-2\,Q^2\,r^2\,\sin ^2\theta
\,d\varphi ^2,
\label{9}
\end{equation}
where $E=1-\frac{4\,M}r$.

We think that another important and challenging problems related
to these
black hole solutions is that of their energy distributions. We
consider that
a good manner to evaluate the energy associated with a black hole
solution
in string theory in the one which used the M\o ller
energy-momentum complex
[7]. This is because of the above mentioned important results
obtained in
the M\o ller prescription and, also, of the Lessner opinion [10]
and
Cooperstock very important hypothesis [16].

In the following we present the definition of the M\o ller
energy-momentum
complex. Its good properties indicates it as a good tool for
energy-momentum
localization in general relativity.

As we pointed out above, M\o ller gave an expression which allows
us to
perform the calculations in any coordinate system. Also, M\o ller
argued
that his expression enables one to obtain the same values for the
total
energy and momentum as the Einstein energy-momentum complex for a
closed
system.

The M\o ller energy-momentum complex $M_i^{\;k}$ [14] is given by

\begin{equation}
M_i^{\;k}={\frac 1{8\,\pi }\,}\chi _{i\;\;\;,l}^{\;kl}, 
\label{10}
\end{equation}
where

\begin{equation}
\chi _i^{\;kl}=\sqrt{-g}\,\left( {\frac{\partial g_{in}}{\partial
x^m}}-{%
\frac{\partial g_{im}}{\partial x^n}}\right) \,g^{km}\,g^{nl}. 
\label{11}
\end{equation}

The quantity $\chi _i^{\;kl}$ is the M\o ller superpotential and
satisfies
the antisymmetric property

\begin{equation}
\chi _i^{\;kl}=-\chi _i^{\;lk}.  \label{12}
\end{equation}

$M_0^{\;\,0}$ is the energy density and $M_\alpha ^{\;\,0}$ are
the momentum
density components.

Also, $M_i^{\;k}$ satisfies the local conservations laws

\begin{equation}
{\frac{\partial M_i^{\;k}}{\partial x^k}}=0.  \label{13}
\end{equation}

The energy of the physical system in a four-dimensional
background is given
by

\begin{equation}
\begin{tabular}{c}
$E=\int \hskip-7pt\int \hskip-7pt\int
M_0^{\,\,\,0}\,dx^1\,dx^2\,dx^3=$ \\
\\
$={\frac 1{8\,\pi }}\int \hskip-7pt\int \hskip-7pt\int
{\frac{\partial \chi
_0^{\,\,0l}}{\partial x^l}\,}dx^1\,\,dx^2\,\,dx^3.$%
\end{tabular}
\label{14}
\end{equation}

Using the Gauss theorem we get the energy contained in a sphere
of radius $r$%
. We have

\begin{equation}
E=\frac 1{8\,\pi }\oint \chi _0^{\;0l}\,\,d\theta \,d\varphi . 
\label{15}
\end{equation}

First, we consider the dual solution in the string frame which is
the
magnetic black hole described by the metric given by the (6). For
this
solution the non-zero $\chi _i^{\;kl}$ components of the M\o ller
energy-momentum complex are given by

\begin{equation}
\chi _0^{\;01}=\,\frac{2\,M^2-Q^2}{M\,r-Q^2}r\,\sin \theta
\label{16}
\end{equation}

Now, we use (15) and (16) and we obtain the for the energy
distribution the
expression

\begin{equation}
E(r)=\frac 12\frac{2\,M^2-Q^2}{M\,r-Q^2}r.  \label{17}
\end{equation}

In the Fig.1 we give the graphic representation of the energy.
$E$ is
plotted against $r$ on X-axis and $Q$ on Y-axis. For the mass $M$
we get the
value $M=1$.
\begin{figure}[h!tbp]
\centering
\includegraphics[width=\columnwidth]{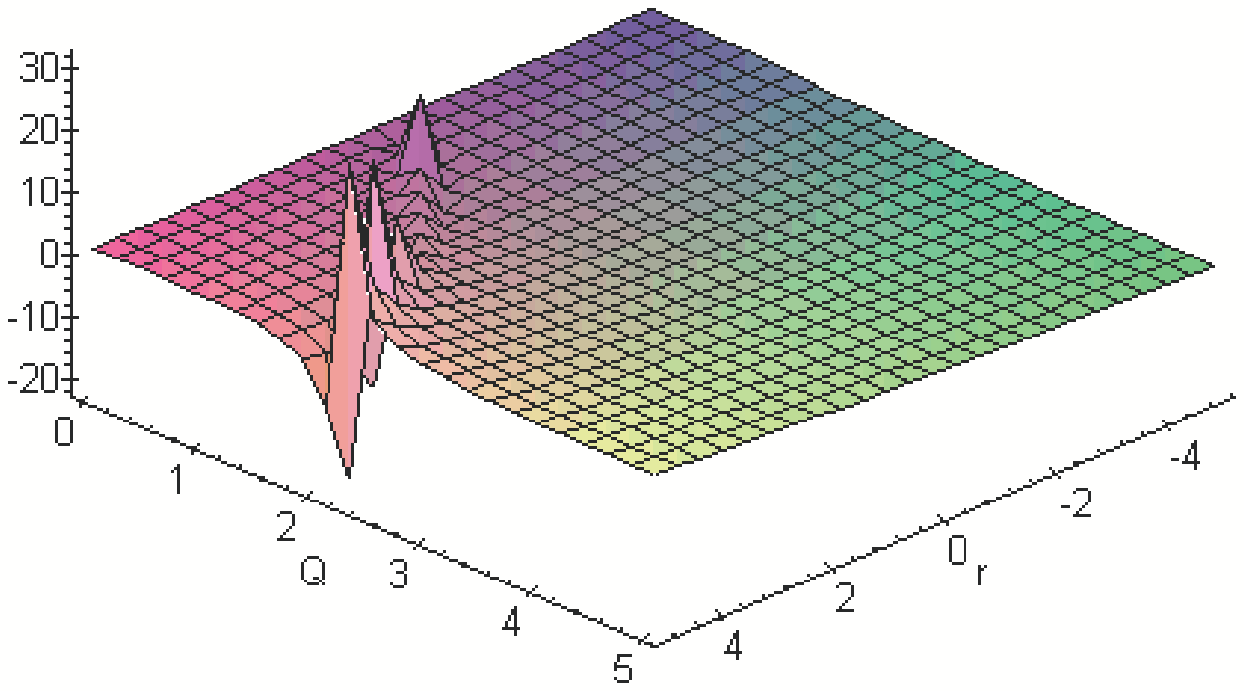}
\caption{}
\end{figure}

From (17) it results that the energy depends on the mass $M$ and
charge $Q$.

For the solution described by the metric given by (8) we get the
non-zero $%
\chi _i^{\;kl}$ components of the M\o ller energy-momentum
complex

\begin{equation}
\chi _0^{\;01}=(\frac{2\,r^2}{\gamma
^2}-\frac{2\,\sqrt{2}\,M\,r}Q)\sin
\theta \,.  \label{18}
\end{equation}

With (15) and (18) we have for the energy

\begin{equation}
E(r)=(\frac{r^2}{\gamma ^2}-\frac{\sqrt{2}\,M\,r}Q).  \label{19}
\end{equation}

In the Fig.2 $E$ is plotted against $r$ on X-axis and $Q$ on
Y-axis. For the
mass $M$ and $\gamma $ we get the values $M=1$ and $\gamma =1$.
\begin{figure}[h!tbp]
\centering
\includegraphics[width=\columnwidth]{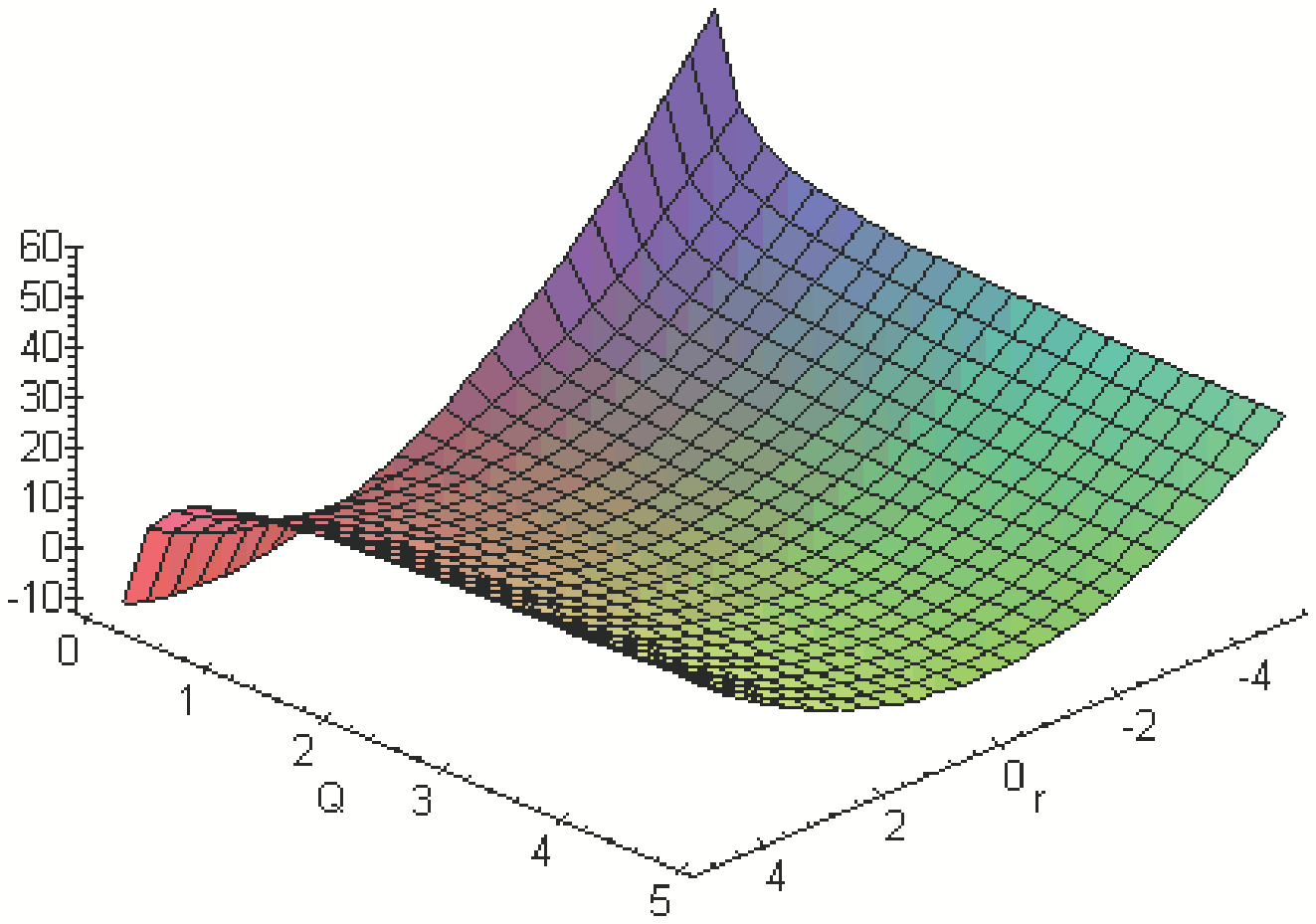}
\caption{}
\end{figure}

We observe from (19) that the energy distribution depends on the
mass $M$,
charge $Q$ and $\gamma $.

In the case of the magnetic black hole given by (9) we obtain the
non-vanishing $\chi _i^{\;kl}$ components of the M\o ller
energy-momentum
complex

\begin{equation}
\chi _0^{\;01}=\frac{8\,M\,Q\,^2}{\gamma ^2}r\,\sin \theta \,. 
\label{20}
\end{equation}

Using (15) and (20) we obtain for the energy distribution

\begin{equation}
E(r)=\frac{4\,M\,Q^2\,r}{\gamma ^2}.  \label{21}
\end{equation}

In the Fig.3 we have the graphic representation of the energy.
$E$ is
plotted against $r$ on X-axis and $Q$ on Y-axis. For the mass $M$
and $%
\gamma $ we get the values $M=1$ and $\gamma =1$.
\begin{figure}[h!tbp]
\centering
\includegraphics[width=\columnwidth]{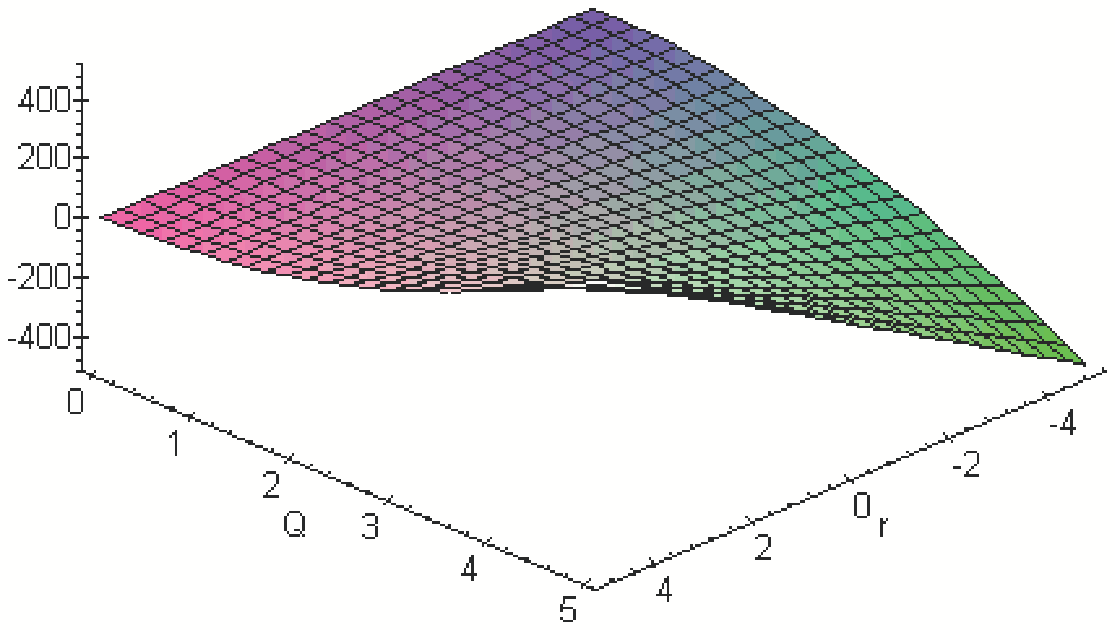}
\caption{}
\end{figure}

From (21) it results that the energy depends on the mass $M$,
charge $Q$ and
$\gamma $.

\section{DISCUSSION}

The subject of the localization of energy continues to be an open
one since
Einstein has given his important result of the special theory of
relativity
that mass is equivalent to energy. Misner et al [17] sustained
that to look
for a local energy-momentum means that is looking for the right
answer to
the wrong question. Also, they concluded that the energy is
localizable only
for spherical systems. On the other hand, Cooperstock and
Sarracino [18]
demonstrated that if the energy is localizable in spherical
systems then it
is also localizable in any space-times. Bondi [19] gave his
opinion that ''a
nonlocalizable form of energy is not admissible in general
relativity,
because any form of energy contributes to gravitation and so its
location
can in principle be found''. Also, Chang, Nester and Chen [20]
showed that
the energy-momentum complexes are actually quasilocal and
legitimate
expression for the energy-momentum. They concluded that there
exist a direct
relationship between energy-momentum complexes and quasilocal
expressions
because every energy-momentum complexes is associated with a
legitimate
Hamiltonian boundary term.

Even the method of localization of energy with several
energy-momentum
complexes has many adepts there was, also, many criticism related
to the use
of the energy-momentum complexes in energy and momentum
localization. The
main lack of the energy-momentum complexes is that most of these
restrict
one to calculate in quasi-Cartesian coordinates.

As we pointed out above, in Introduction, the most
energy-momentum complexes
are dependent on the coordinate system. Only the M\o ller
energy-momentum
complex allows us to make the calculations in any coordinate
system. Another
argument that sustains the use of the M\o ller prescription is
the Lessner
[10] conclusion that the M\o ller energy-momentum complex is a
powerful
representation of energy and momentum in general relativity. The
viewpoint
of Lessner [10] is that ''The energy-momentum four-vector can
only transform
according to special relativity only if it is transformed to a
reference
system with an everywhere constant velocity. This cannot be
achieved by a
global Lorentz transformation''.

In this paper we choose to study the energy associated with some
stringy
black hole solution [13] because those solutions are interesting
to study.
we obtain results for three stringy black hole solutions in the
string
frame. The energy of the magnetic black hole, which is the dual
solution in
the string frame, in eq. (17) is $ \lim_{r\to \infty} E(r)
=M-\frac{Q^2}{2\,M}$.

Our result show that the total energy is dependent on charge $Q$
and differs
to previous investigations. It shows us that the stringy black
holes in
string frame are interesting for study. The total energy of the
other two
non-asymptotically flat black hole solutions, in eqs. (8) and
(9), become
infinite. From eqs. (19) and (21) we found out that the M\o ller
energy-momentum complex furnishes us consistent results of
non-asymptotically flat black hole solutions.

\end{document}